\documentstyle[preprint,aps,graphicx]{revtex}
\tightenlines
\begin{document}
\newcommand{\bref}[1]{eq.~(\ref{#1})}
\newcommand{\be}{\begin{equation}}
\newcommand{\en}{\end{equation}}
\newcommand{\bs}{$\backslash$}
\newcommand{\us}{$\_$}

\title{Diffusers for holographic stereography}
\author{Lars Egil Helseth${}^{1}$, Ingar Singstad${}^2$}
\address{${}^1$ University of Oslo, Department of Physics, N-0316 Oslo, Norway }
\address{${}^2$ University of Bergen, Department of Physics, N-5007 Bergen, Norway }

\maketitle
\begin{abstract} 
Holographic diffusers have long been recognized as versatile 
components with a broad number of applications. In this work we discuss
holographic diffusers for projection of laser light from a Liquid Crystal Display(LCD) onto a 
holographic recording medium. In holographic stereography, projection of the 
information from a LCD onto a holographic recording medium has traditionally 
been done by a lens or a ground glass. It is suggested that the holographic
diffusers can replace these elements and improve image quality and light 
economy. 

\end{abstract}

\narrowtext

\newpage

\section{Introduction}
Storage of data by means of holography has long been an active field of 
research. Usually, the 
storage must be done in coherent light. However, by the
application of holographic stereography, this requirement can be relaxed.    
A holographic stereogram is created by optical multiplexing of a finite number of 
viewpoints onto different regions of a high resolution recording medium
(see ref.\cite{Okada,Klug} and references therein).   
The perspectives(or viewpoints) are usually displayed on a Liquid Crystal
Display(LCD), and then projected onto a small area of the recording medium.  
Traditionally the multiplexing have been done in one of two ways\cite{Okada,Klug};
1)A cylindrical lens is placed behind a LCD, focusing the laser light into a small slit. This kind of 
holographic stereogram is usually refered to as a multiplex stereogram. The cylindrical lens conserve the image plane 
characteristics in the vertical direction, and gives rise to a rainbow colored 3D 
image\cite{Okada}. Unfortunately, the intensity distribution at the focal plane 
of the cylindrical lens usually contains large variations due to diffraction. 
Unless the recording media has large dynamic range, these intensity variations 
gives rise to saturation and loss of information\cite{Klug}.
2)The alternative is to produce a socalled DeBitetto stereogram\cite{Klug,DeBitetto}.
In this case an ordinary diffuser is placed right behind the LCD, and diffuses the light 
in both the horizontal and vertical direction. 
Since the slit width is of order 1mm, most of the light from the diffuser is 
scattered outside the slit, giving very long exposure times. To reduce the 
exposure times, a ground glass with finite diffusing angle is applied. 
Unfortunately, this results in an uneven illuminated 3D image, 
especially when the ground glass is close to the holographic film. 
Note also that a ground glass will diffuse the light in the vertical direction as well, thus destroying the image plane characteristics. The DeBitetto stereogram is 
therefore not suitable for production of white light viewable transmission 
stereograms on thin emulsions. However, the image may be transferred to the 
plane of a second filmplate. Such image plane stereograms can be displayed in 
white light with very little color blur\cite{Klug}.

Ideally, the intensity distribution at the emulsion plane should be as rectangular as possible, 
it should contain equal contributions from all parts of the LCD(to obtain an 
even illuminated 3D image), and all the incoming light should be utilized. 
Furthermore, if the stereogram is recorded on a thin 
silver halide emulsion, the image plane characteristics must be preserved, to 
suppress cross-talk images\cite{Klug}. None of the traditional techniques 
fulfill all these requirements. Therefore, alternative solutions must be found.

Pseudorandom diffusers seems to be a mature solution for both 
volume data storage and holographic stereograms\cite{Nakayama,Yamaguchi,Gao},
due to their ability to distribute the light at the focal plane, without destroying 
the image plane characteristics.
Pseudorandom diffusers were introduced in holographic stereography by Klug $et$
$al$\cite{Klug}, and independently by Yamaguchi $et$ $al$\cite{Yamaguchi}, and 
were shown to increase the image quality of the holographic stereograms.     
   
In this paper we propose two different holographic diffusers. The first type, which
we will simply call a 'holographic diffuser', can be considered as an alternative to the classical
diffuser in a DeBitetto stereogram. It is capable of directing more light
into the slit in the holoprinter, and perhaps also improve the image quality,
by providing a more even illuminated image than that resulting from a ground 
glass. The second type, which we will call a 'multiple-point holographic diffuser', can
be considered as an alternative to the cylindrical lens in a multiplex
stereograms. We think it is capable of improving the image quality of multiplex 
stereograms, without significant reduction of the intensity incident on the 
recording medium. To confirm the quality of the holographic diffusers described above, 
we recorded a few holographic stereograms with the setup shown in fig. \ref{f6},
which will be described later.
This system will be referred to as the holoprinter. 
We would like to point out that the diffusers should be designed according to standard
stereographic projection relationships, to ensure that the final 
images are free of anamorphic distortions(see e.g. \cite{Okada,Klug,DeBitetto} 
and references therein).

\section{Holographic diffuser}
As mentioned in the previous section, application of a ground glass results 
in loss of light and uneven illumination of the different parts of the 3D image. 
A solution to this problem may be holographic diffusers. 
Holographic diffusers have been developed for many display 
purposes\cite{Pawluczyk}. They are 
superior to classical diffusers in that they can redirect and control up to 
$100\%$ of the incoming light. A holographic diffuser suitable for holographic 
stereography can be created as follows: 
1)First a strip diffuser is created by covering an isotropic diffuser with a 
slit.  
2)Next a hologram is recorded with this strip diffuser as the object. A hologram recorded in this way will simply 
be called a $holographic$ $diffuser$. 

To use this holographic diffuser in the production of a holographic
stereogram, it must be placed in contact with the LCD screen. As pointed out 
in the introduction, the purpose of the LCD is to provide a perspective picture 
which can be projected onto a small area of the holographic film. 
By illuminating the LCD(+holographic diffuser) with a proper reconstruction 
beam, the real image of the strip diffuser is projected 
into the slit of the holoprinter. We believe that the real image of the
slit should match that of the holoprinter since this ensures 
that the whole perspective picture is passed through the slit in the 
holoprinter, and that maximum amount of light is utilized. If the slit width of
the holographic difuser is too small, we probably get picket fence 
artifacts(a detailed description of this problem is given in ref.\cite{Klug}). If it is too big, some of the perspective 
picture is cut off, and there is also some loss of light.     

To produce a holographic diffuser using the principles described 
above, we covered a diffuser(opal diffuser) with a 0.5mm slit. This strip diffuser was placed 
slightly away from focus of a cylindrical lens, see fig. \ref{f1}. The lens was 
introduced to increase the amount of light passing through the strip diffuser, without introducing too
large speckles. Since the lens was placed in front of the diffuser, its quality 
was not important. A collimated reference beam was illuminating the film at an angle of 
$\sim 50^{\circ}$, and the distance between the diffuser and the holographic 
films was approximately 30cm. With this setup we recorded a hologram with a 20mW HeNe laser 
on BB640 thin silver halide emulsions from 
HRT technologies. The size of the exposed film was 12.7$\times$10cm, and we used
a collimated reference beam which was larger than the film(and therefore
considered uniform). The film was processed by a procedure similar to that 
suggested by Belendez et.al.\cite{Belendez}, resulting in Silver Halide 
Sensitized Gelatin(SHSG). In this way we were able to produce a holographic
diffuser with little unwanted scattering and $40\%$ diffraction 
efficiency. During reconstruction, the real image will form a rectangular 
strip of 0.5mm width. We found that the effective(actual) size of the diffuser was 
10$\times$10cm(the holographic film was partially screened by the holder). 
Since the distance between the original strip diffuser and the holographic film
was 30cm, this results in a convergence angle of approximately 
$\sim 9.5^{\circ}$ during reconstruction.

\section{Multiple-point holographic diffuser}
In many applications we are interested in conserving the image plane
characteristics in the vertical direction, to be able to display the (thin
emulsion) stereogram in white light\cite{Okada}. An optical element which does this 
can also be created by application of holographic principles. 
The experimental procedure is illustrated in fig. \ref{f2}.
We recorded a DeBitetto stereogram of a rectangular 'point' diffuser. 
The 'point' was a Edmund Opal diffuser of size 1mm$\times$1.2mm. 120 element 
holograms were recorded on 12.7$\times$10cm BB520 with a 50 mW 532 nm 
$Nd:YVO_{4}$ laser. The BB520 film was translated by 0.9mm between each 
exposure, equal to the width of the the slit covering the filmplate. 
The distance between the 'point' diffuser and the holographic film 
was approximately 30cm. The effective size of the diffuser was found to be 
10.8$\times$9cm, which results in a convergence angle of 
approximately $\sim 8.5^{\circ}$ during reconstruction.  
A reference beam illuminated the film from above with an angle $\sim 50^{\circ}$. 
This reference beam could be considered uniform, since it was much larger than the
rectangular slit in the holoprinter. The BB520-film was processed by a procedure similar to that 
suggested by Belendez et.al.\cite{Belendez}, resulting in Silver Halide 
Sensitized Gelatin(SHSG). In this way we were able to produce a 
diffuser with little unwanted scattering and $30\%$ diffraction 
efficiency. When the real image of this multiple element 
hologram was reconstructed, the diffuse points were reconstructed side by side, 
thus creating a vertical line, see fig. \ref{f2}. We will name this kind of 
hologram a $multiple-point$ $holographic$ $diffuser$, since it acts differently 
from the holographic diffuser discussed in the previous section, and is supposed
to be used during production of a multiplex stereogram. 
By illuminating the LCD(+ multiple-point holographic diffuser), the perspective 
information on the LCD is projected onto a line of size $\sim 1.2mm$. 
The line looks continous, since the width of each point(1mm) is slightly 
larger than the distance translated between each exposure(0.9mm). 
This gives a small overlap between reconstructed 'points'. Unfortunately, we
have not conducted any analysis to see how this overlap influences the image
quality. More research is needed to find out how this as well as the size of the slit
influences the final image.

\section{Measurements of the intensity distributions}
We have measured the intensity distributions from the two holographic diffusers 
described above. The experimental setup used is shown in fig. \ref{f3}. A detector from Pasco 
Scientific(model 8020) was covered with a small vertical slit of size $10\mu m$. 
The detector can be moved(longitudinally) to the proper 'focal' plane of the diffuser 
by a micrometer screw. A second micrometer screw was used to move the 
detector(transversally) across the intensity distribution in steps of $10\mu m$. 
The inaccuracy of the translating mechanism is about 5$\mu m$. 
The same collimating lens was used during recording and reconstruction, and the 
collimated reference beam was larger than the exposed 
film(12.7$\times$10cm), and therefore considered uniform. The detector 
was placed at the point were the real image was supposed to be located. 
In both cases this distance was 30cm(as indicated in the descriptions above). 
By translating the detector along the optical axis, we 
measured the highest possible peak intensity 30cm from the diffusers 
within an accuracy of $\pm 1cm$.       
 
Fig. \ref{f4} shows the relative intensity distribution of 
the holographic diffuser. The width of the distribution is 0.5mm, which is the same as the 
width of the strip diffuser used to record the holographic diffuser. Some light 
is distributed outside the original boundaries, which may be caused by diffraction 
from the edge, unwanted scattering, or misalignment during reconstruction. 
The intensity variations within the passband is about 0.3. A major contributor 
to the rapid intensity variations is speckle noise 
introduced by the ground glass, but the emulsion will also contribute. Note also that the intensity distribution is 
not completely symmetric. We believe this is due to a slight misaligment 
of the slit during recording. 
In fig. \ref{f5}, the intensity distribution at the 'focal' point of the 
multiple-point diffuser is shown. Note that the 
measured intensity distribution from the multiple-point diffuser is nearly 
rectangular. Again, some light is distributed outside the 
boundaries of the original 'point' diffuser(width 1.2mm). 
It is seen that the intensity variations within the passband is about 0.5. 
Furthermore, the variations are much more rapid than in fig. \ref{f4}. 
This is believed to be due to the fact that we used different
emulsions(BB520 and BB640), although this has not been completely 
clarified.

\section{Setup for recording holographic stereograms}
To confirm the quality of the holographic diffusers described above, we 
recorded a few holographic stereograms with the setup shown in fig. \ref{f6}. 
The laser is a 20mW 632nm HeNe laser or a 50mW 532nm $Nd:YVO_{4}$-laser. 
The reference and object beams are separated by a beamsplitter, and then 
transported to different levels. The object beam is cleaned and expanded by a spatial filter, after which it  
illuminates the LCD. The LCD is placed in close contact with the 
diffuser, which directs the light into the slit of the printer. 
The reference beam is also cleaned and expanded, and then directed 
towards the slit in the printer. Each perspective picture on the LCD is 
exposed onto a strip of the holographic film, one by one. In total, a 
holographic stereogram is generated. We applied both the holographic and 
multiple-point diffusers in front of the LCD, with good results.     
We hope to publish a more detailed evaluation of the holographic stereograms 
produced with this holoprinter in a separate paper.

In conclusion, we have described two different diffusers which may be of 
interest in holographic stereography. We have performed some simple
measurements to get a better picture of the performance of these diffusers.
However, many questions are still not answered, and should be the subject of 
future investigations. In particular it would be interesting to see 
the influence of different numerical apertures, the reference beam angle, the 
polarization of the incident beam, and the size of the slit. Also the influence
of overlap in the multiple-point diffuser should be studied in more detail.

This work was done while L. E. Helseth was at the University of Bergen.

\newpage
\begin{figure}
\caption{ Recording and reconstruction of the holographic diffuser described in
the text. An ordinary diffuser is covered by a small slit, and is 
illuminated from behind. The diffused light interfere with the reference beam at the 
filmplane. When reconstructed, the real image is a strip of the same size as 
the original strip diffuser.  
\label{f1}}
\end{figure}

\begin{figure}
\caption{ Recording and reconstruction of the multippel-point holographic
diffuser described in the text. The light from a point diffuser is  applied as
object beam. Between each exposure the film is transported a distance $s$. 
\label{f2}}
\end{figure}

\begin{figure}
\caption{ Setup for measuring the intensity distributions. A detector is 
covered by a $10\mu m$ slit. During measurement, the detector is translated 
in steps of $10\mu m$.
\label{f3}}
\end{figure}

\begin{figure}
\caption{Intensity distribution from the holographic diffuser. The diffuser 
is recorded on BB640, and reconstructed with a 20mW $632nm$ HeNe laser.     
\label{f4}}
\end{figure}

\begin{figure}
\caption{Intensity distribution from the multippel-point holographic diffuser. 
It was recorded on BB520, and reconstructed with a 50mW $Nd:YVO_{4}$ laser. 
\label{f5}}
\end{figure}

\begin{figure}
\caption{Setup for recording a holographic stereogram with a holographic 
diffuser or multiple-point holographic diffuser. The dashed line represents 
the beampath on the second level. The reference beam illuminate the filmplate 
from above. 
\label{f6}}
\end{figure}

\newpage
\centerline{\includegraphics[width=14cm]{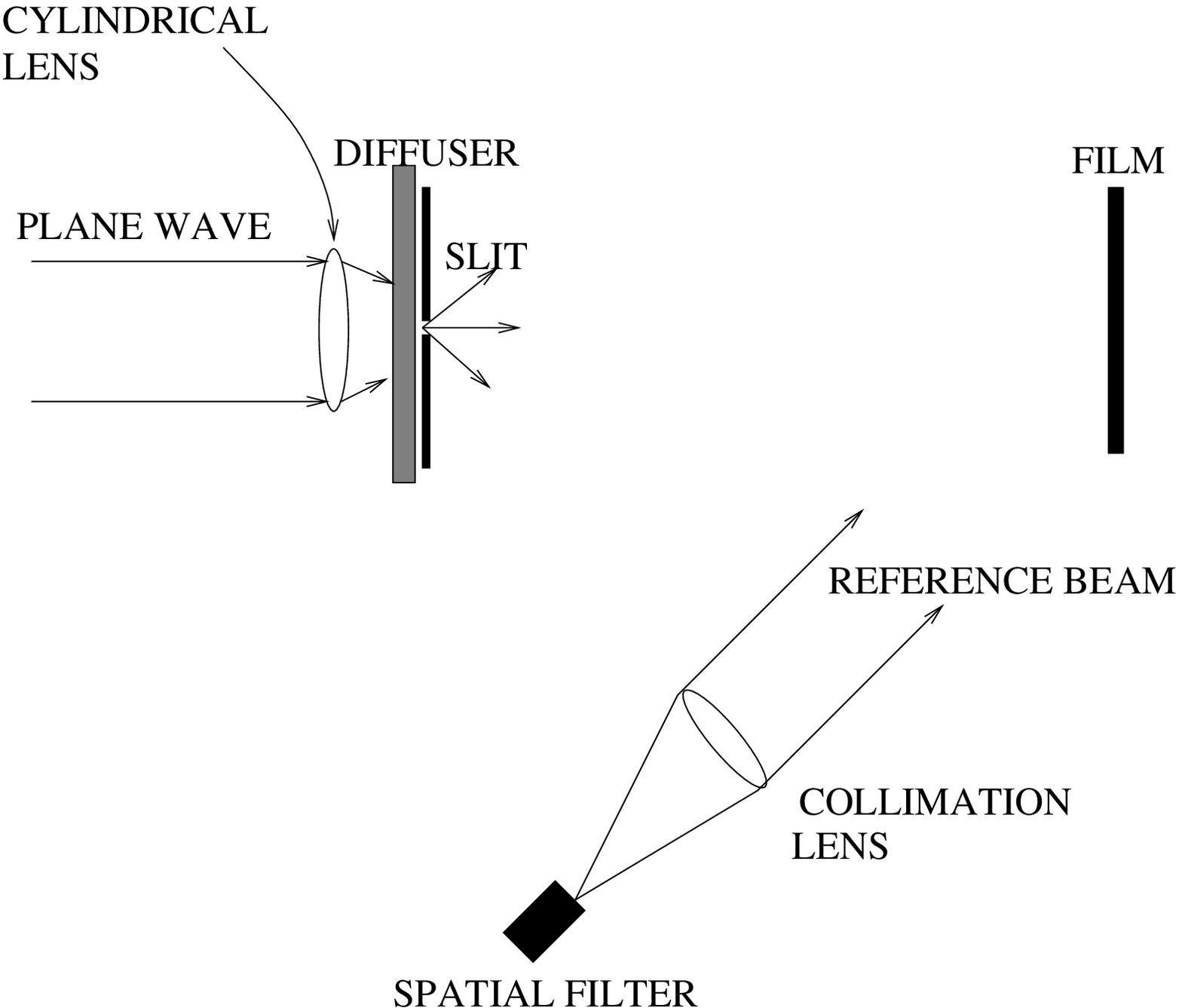}}
\vspace{2cm}
\centerline{Figure~\ref{f1}}

\newpage
\centerline{\includegraphics[width=14cm]{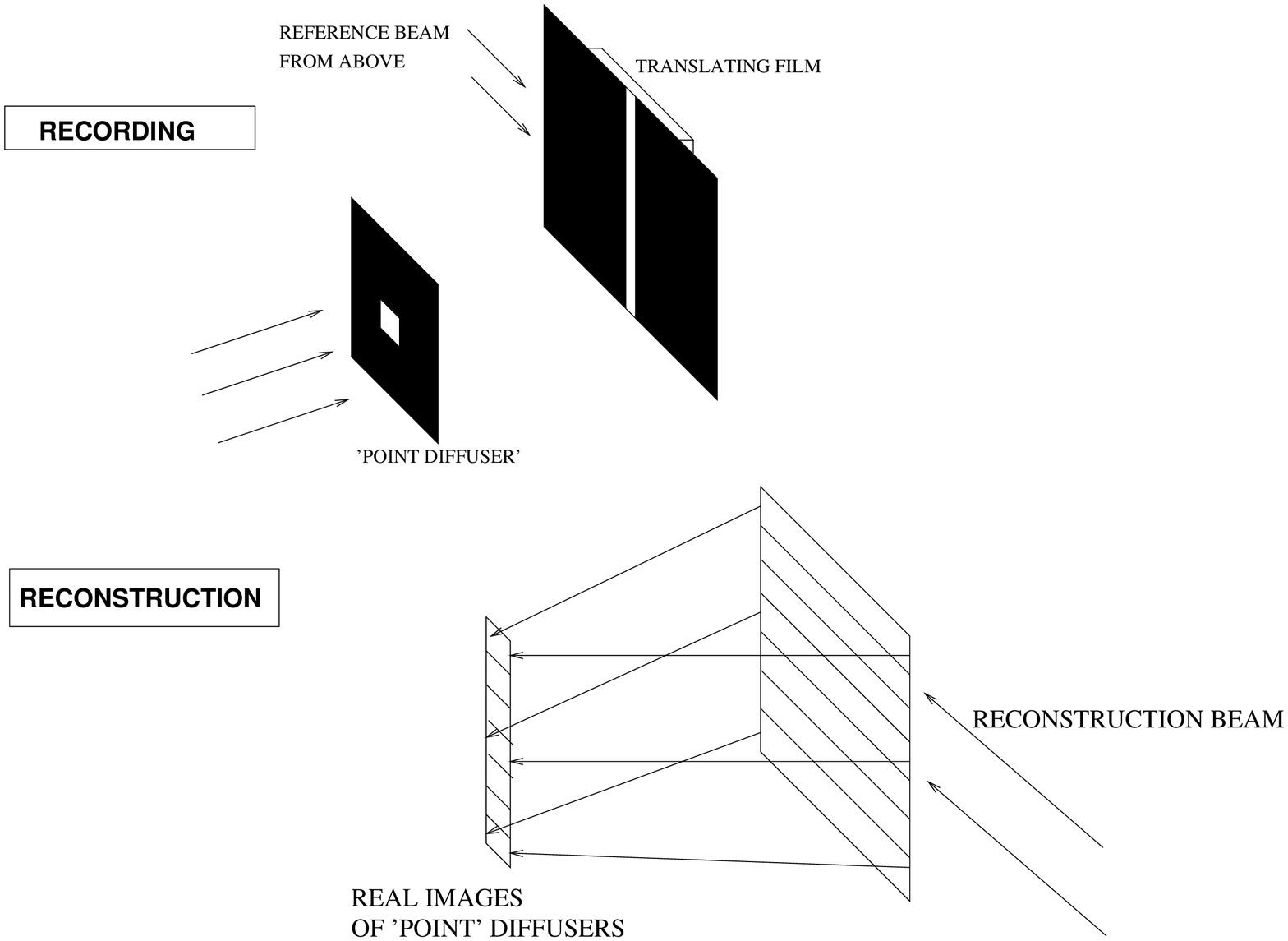}}
\vspace{2cm}
\centerline{Figure~\ref{f2}}

\newpage
\centerline{\includegraphics[width=14cm]{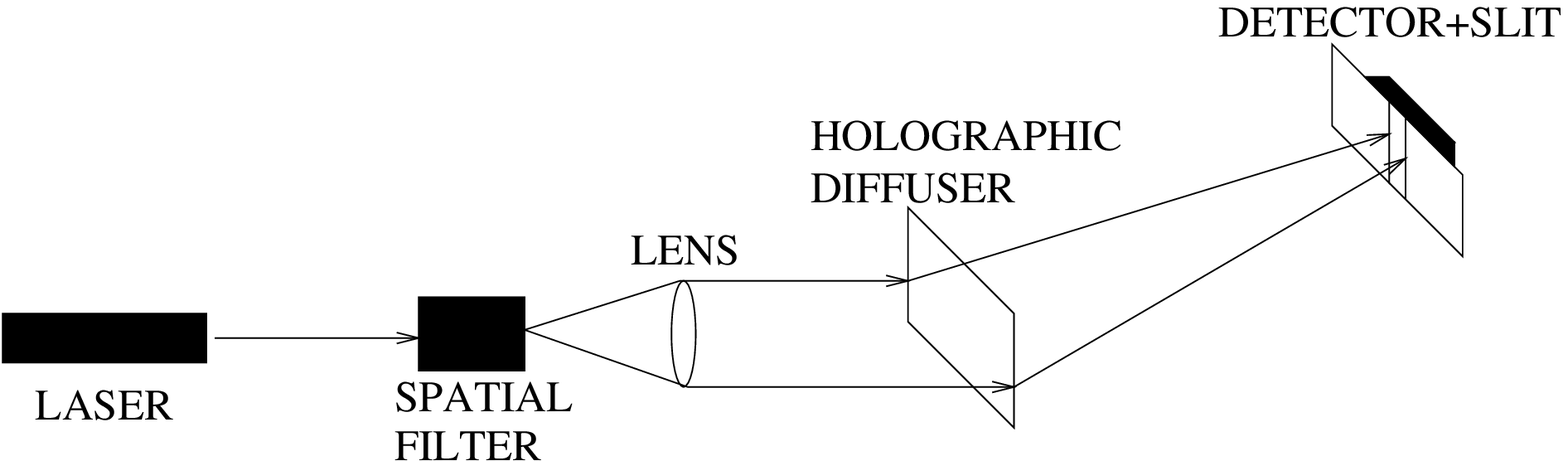}}
\vspace{2cm}
\centerline{Figure~\ref{f3}}

\newpage
\centerline{\includegraphics[width=14cm]{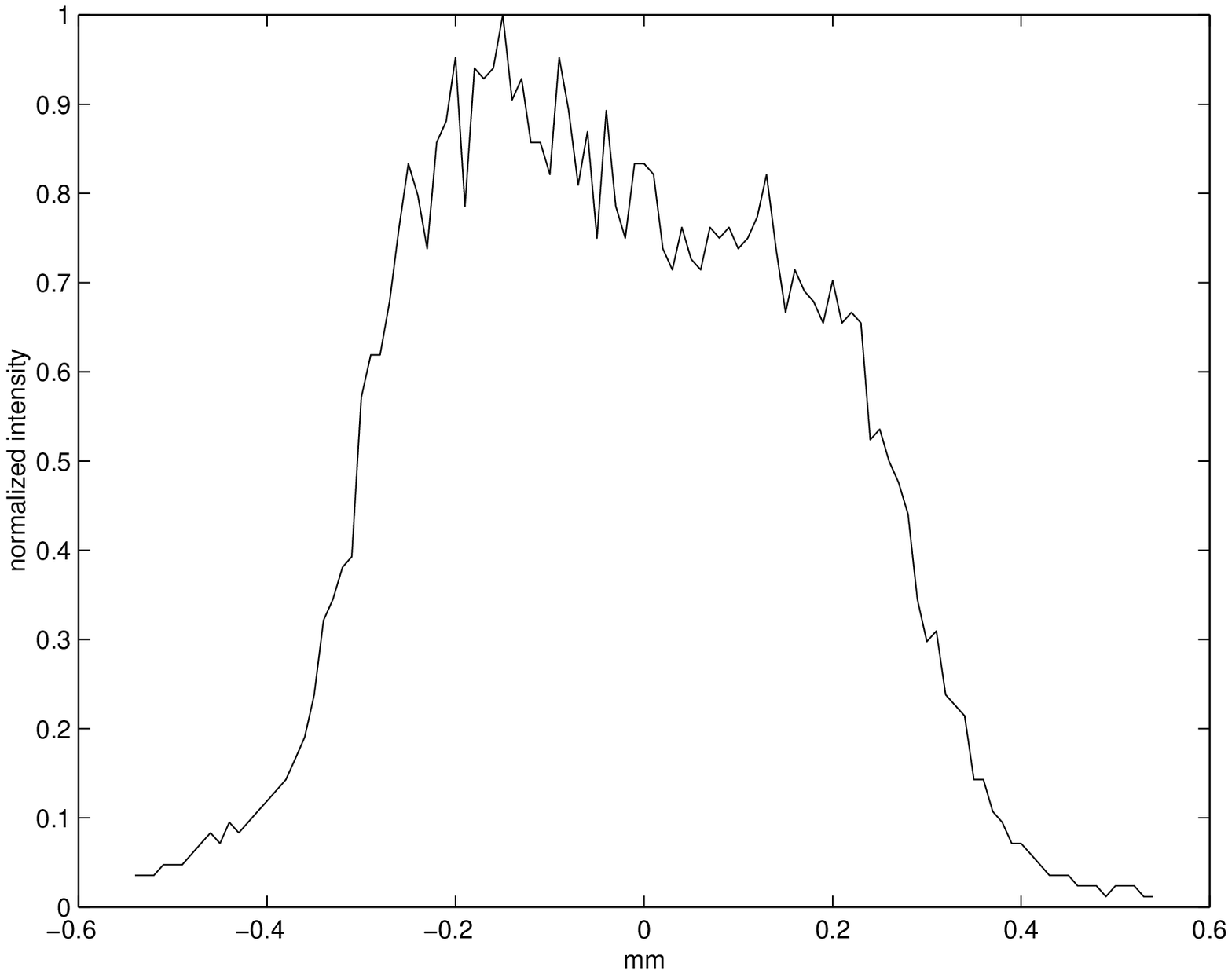}}
\vspace{2cm}
\centerline{Figure~\ref{f4}}

\newpage
\centerline{\includegraphics[width=14cm]{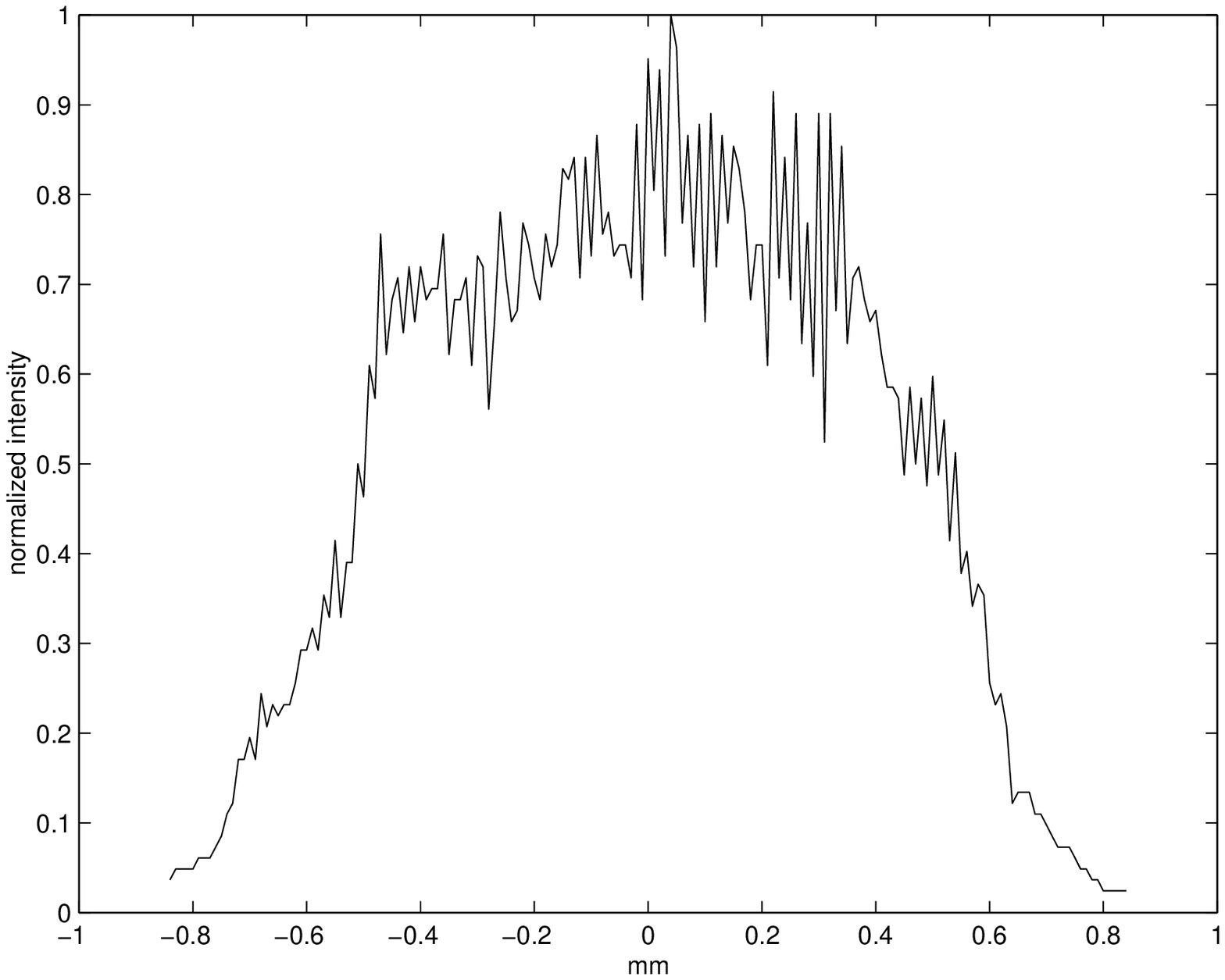}}
\vspace{2cm}
\centerline{Figure~\ref{f5}}

\newpage
\centerline{\includegraphics[width=14cm]{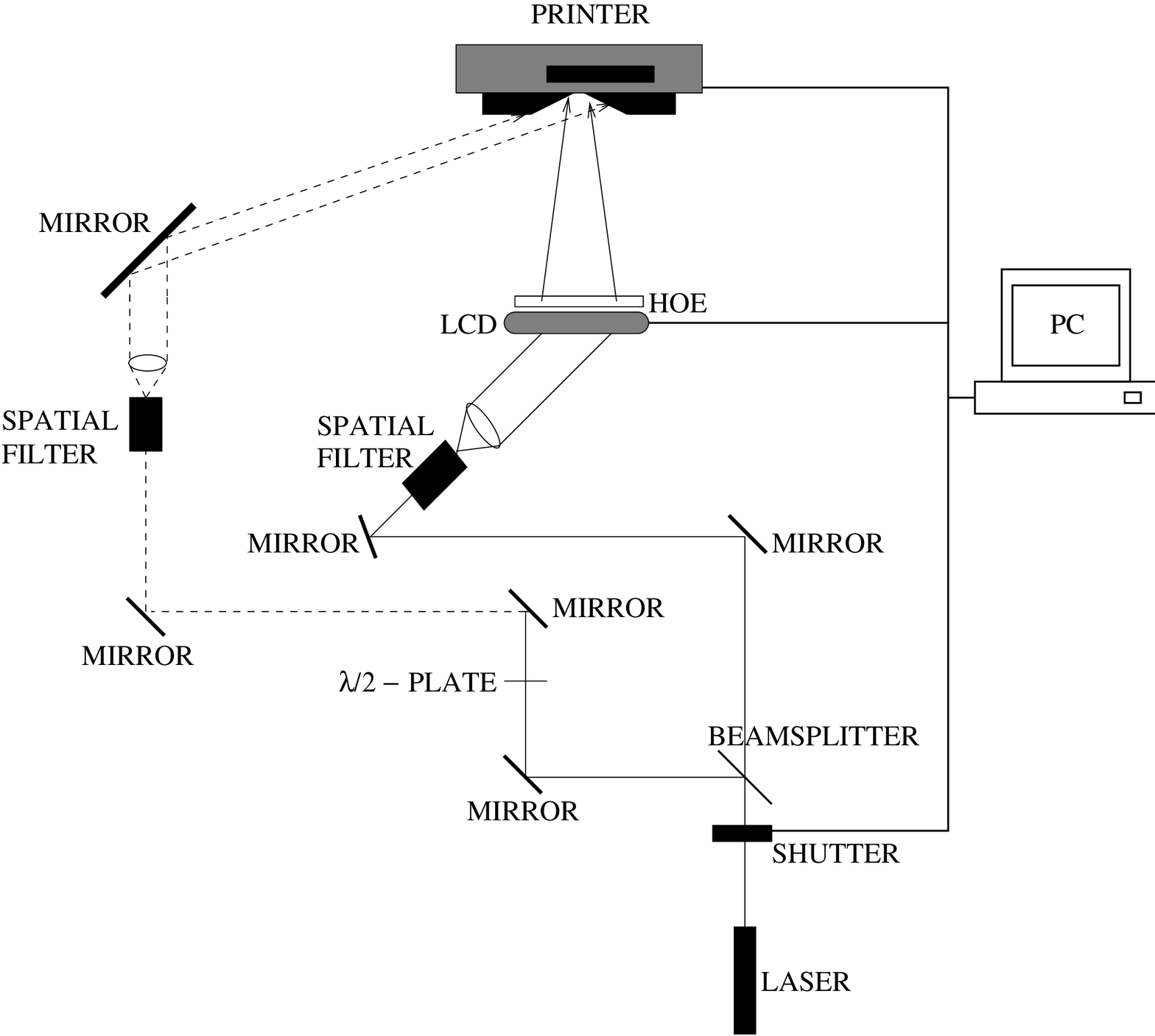}}
\vspace{2cm}
\centerline{Figure~\ref{f6}}

\end{document}